\documentclass[12pt,preprint]{aastex}
\def\msun{$M_{\odot}$}

\usepackage{fullpage}
\begin{document}

\thispagestyle{empty}
\vspace*{2cm}

\begin{center}

\Large
{\bf Measuring the Spins of Stellar-Mass Black Holes} \\
\normalsize

\vspace{10mm}

{\large Jeffrey E. McClintock} \\
\vspace{1.0mm}
{\it Harvard-Smithsonian Center for Astrophysics, 60 Garden Street,
Cambridge MA 02138} \\

\vspace{3mm}

{\large Ronald A. Remillard} \\
\vspace{1.0mm}
{\it Kavli Institute for Astrophysics and Space Research, Massachusetts
Institute of Technology, Cambridge, MA 02139} \\

\vspace{15mm}

Science white paper prepared for: \\
\vspace{2mm}
\noindent The Astronomy \& Astrophysics Decadal Survey Committee \\
\vspace{2mm}
\noindent The Science Frontier Panel on Stars and Stellar Evolution
(SSE) \\
\vspace{2mm}
\noindent The Science Frontier Panel on Cosmology and Fundamental
Physics (CFP) \\

\vspace{5mm}

February 2009 \\

\end{center}

\newpage

\setcounter{page}{1}

\section{Introduction and Scientific Context}

``If the cosmological problem is the number one problem of astronomy,
then problem number two should be the problem of black holes$^1$.''
These words of Nobelist V. L. Ginzburg from the 1980s ring true today.
Furthermore, the two problems are inextricably linked because the
nascent Universe can only be compared to a black hole (BH).  In
addition, the evolution of the Universe is profoundly affected by the
presence of BHs.  For example, BHs power AGN, which interact with and
greatly modify their host galaxies and clusters; and they are an
ultimate endpoint of stellar evolution, with one percent or so of the
Milky Way's baryons having already been swept away into stellar-mass
BHs.

With no useful theory of quantum gravity on the horizon, fundamental
physics is stymied.  In this instance, astronomy can serve physics by
prosecuting the study of astronomical BHs, possibly the only kind of BH
that we will ever know.  In doing so, it is wise to take our cue from
the study of cosmology where no stone is left unturned, where the study
of the CMB, supernovae, galaxy clusters, GRBs, quasars, galactic and
stellar evolution, etc., all play important roles in digging deep.  For
BHs likewise, it makes sense to pursue with equal vigor both
supermassive and stellar-mass BHs, while seeking dynamical evidence for
intermediate-mass BHs.

And it is equally important to use all available data channels.  We can
reasonably expect that LIGO and LISA will provide us with intimate
knowledge concerning BHs.  However, gravitational wave detectors are
unlikely to tell us much about MHD accretion flows in strong fields or
the origin of relativistic jets or about relativistically-broadened Fe
lines and high-frequency quasi-periodic oscillations, phenomena that are
now routinely observed for BHs$^2$.  In short, observations of {\it
accreting} BHs show us uniquely how a BH interacts with its environment.
It behooves us to explore widely because, as in cosmology, it is the
synergistic exploration of all paths that enlightens.  Therefore, it is
important to maintain balance between gravitational-wave studies of BHs
in vacuum and electromagnetic studies of BHs that are situated in
accretion flows.

Today, on the one hand we have solid dynamical evidence for objects of
extraordinary density -- objects such as the two-dozen BHs in X-ray
binaries and SgrA$^*$.  And on the other hand, we have General
Relativity which firmly predicts that these objects have undergone
complete gravitational collapse.  Whether the collapse leads to a
Planck-scale singularity or is mediated by quantum effects on a larger
scale is at present not a practical concern for astronomers.  Our
position is rather the following: While keeping an eye on exotic
physics, we assume that GR is the correct theory of strong gravity, and
we use this venerable theory to interpret our observations of BHs while
searching for inconsistencies and contradictions.  Meanwhile, we note
that there is compelling evidence that stellar-mass BHs in X-ray
binaries and the supermassive BH in Sgr A* have event horizons$^{3}$,
confirming our belief that GR is the correct theory of strong gravity.

Stellar-mass BHs and the measurement of their spin are the subject of
this white paper.  Twenty-two dynamically-confirmed BHs are now
known$^{2,4,5}$.  For a catalog of their host binaries and a schematic
sketch to scale of most of them, see respectively Table 1 and Figure~1
in ref.\ 2.  Seventeen of these BHs are in transient systems and five
are in systems that are persistently X-ray bright (e.g., Cyg X-1).  The
masses of these BHs range from 5--20\msun~with a typical value of
10\msun.  Eighteen are located in the Milky Way, two in the LMC, and two
in other local group galaxies.

\vspace{-5mm}

\section{The Measurement of Spin -- A Frontier in Black Hole Research}

Astrophysical BHs are completely described by the two numbers that
specify their mass and spin$^6$.  BH spin is commonly expressed in terms
of the dimensionless quantity $a_* \equiv cJ/GM^2$ with $|a_*| \le 1$,
where $M$ and $J$ are respectively the BH mass and angular momentum.
While mass measurements of stellar BHs have been made for decades, the
first spin measurements have been achieved only during the past three
years$^{7-14}$.  Meanwhile, the spin of a supermassive BH has also been
measured$^{15,16}$.

Knowledge of BH spin is crucial for answering many key questions, for
example: (1) Are relativistic jets powered by spin?  It is widely
speculated that these jets, observed for at least eight BH microquasars
and hundreds of AGN, are powered by BH spin via a magnetic Penrose
process$^{17}$.  With many secure measurements of spin and mass in hand
it will be possible to attack the jet/spin/Penrose-process connection in
earnest.  (2) What role does spin play in powering a gamma-ray burst?
For example, a great uncertainty in modeling long GRBs is whether one
can arrive at the core-collapse stage with sufficient angular momentum
to make a disk around a BH$^{18}$.  (3) What constraints can be placed
on models of supernovae, BH formation, and BH binary evolution with both
mass and spin in hand$^{19,20}$?  (4) What distribution of BH spins
should LIGO waveform modelers be considering$^{21}$?  (5) For
supermassive BHs, is the distribution of spins of the merging partners
consistent with hierarchical models for their growth$^{22}$?

In this section, we first consider the two techniques that are currently
delivering measurements of spin, namely fitting the thermal X-ray
continuum$^{23}$ and modeling the profile of the Fe K line$^{24}$.
Because spin is such a critical parameter it is important to measure it
by both methods, as this will arguably provide the best possible check
on our results.  {\it Since the continuum-fitting (CF) method cannot be
  applied to AGN, BH binaries are the crucial common ground where both
  current methodologies for measuring spin are now being readily
  applied.}  Secondly, we consider a highly promising and independent
avenue to spin -- high-frequency QPOs.  Finally, we examine X-ray
polarimetry, which has the potential to secure the measurement of spin
via the CF and Fe K methods, while possibly opening a fourth avenue to
spin.

\vspace{-5mm}

\subsection{Current Approaches: The X-ray Continuum and the Fe K Line}

BH spin is measured by estimating the inner radius of the accretion disk
$R_{\rm in}$, which is identified with the radius of the innermost
stable circular orbit $R_{\rm ISCO}$ predicted by GR$^{6}$.  Strong
support for linking $R_{\rm in}$ to $R_{\rm ISCO}$ is provided by
decades of empirical evidence that $R_{\rm in}$ is constant in
disk-dominated states of BH binaries$^{25}$ and by recent MHD
simulations of thin accretion disks$^{26,27}$.  $R_{\rm ISCO}/M$ is a
monotonic function of $a_*$, decreasing from $6GM/c^2$ to $GM/c^2$ as
spin increases from $a_*=0$ to $a_*=1$ (ref.\ 6).  {\it This
relationship between $a_*$ and $R_{\rm ISCO}$ is the foundation of both
the CF and the Fe K methods of measuring spin.}

\begin{figure}
\begin{center}
\includegraphics[width=0.70\linewidth]{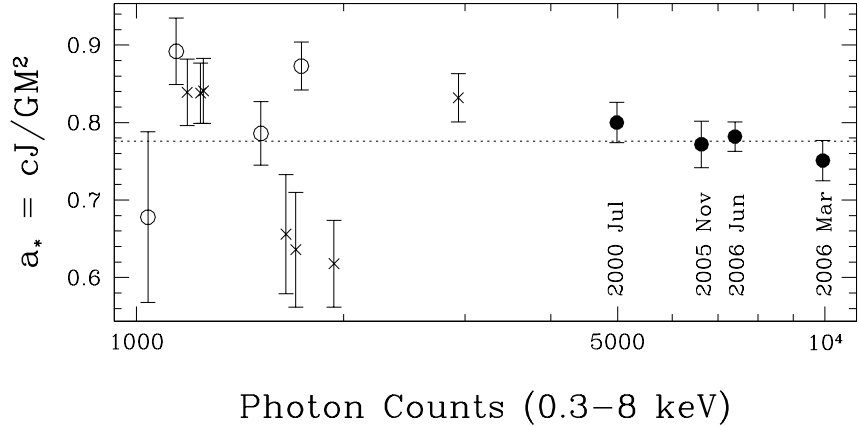}
\end{center}
\vspace{-8mm}
\caption{ Spin results for the BH primary in M33 X-7 obtained by fitting
  {\it Chandra} spectra (filled/open circles) and {\it XMM} spectra
  (crosses) to the relativistic disk model {\sc kerrbb2} (ref.\ 10); the
  data are ordered by total counts.  The four ``gold'' {\it Chandra}
  spectra with $\gtrsim 5000$ counts each (filled circles) yield spin
  estimates that agree with the mean value (dotted line) to within their
  $\approx 2$\% statistical uncertainties, which is remarkable stability
  given that the observations span years (see dates).  Meanwhile, this
  mean value agrees with the mean spin for the 11 low-quality spectra
  with $<3000$ counts to within $\approx 1$\%.  Including all
  observational uncertainties (e.g., BH mass), one obtains $a_* = 0.77
  \pm 0.05$.}
\end{figure}

\begin{figure}
\begin{center}
\includegraphics[width=0.70\linewidth]{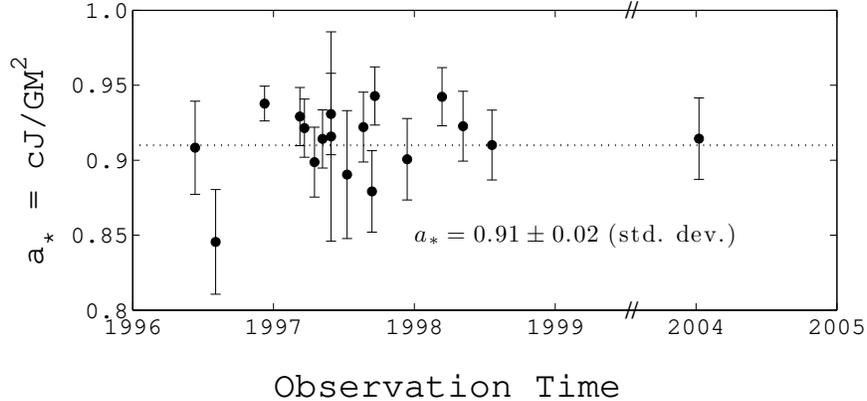}
\end{center}
\vspace{-5mm}
\caption{ Spin results versus time for the BH primary in LMC X-1
  obtained by fitting {\it RXTE} spectra to the relativistic disk model
  {\sc kerrbb2} (ref.\ 11).  The spectra were selected as minimally
  Comptonized from a complete sample of 55 {\it RXTE} spectra.  As
  indicated, the scatter about the mean value of $a_*$ is small.
  Including all model-parameter and observational uncertainties (e.g.,
  $\alpha$-viscosity and BH mass), one obtains $a_* =
  0.90_{-0.09}^{+0.04}$.  Virtues of {\it RXTE} are its good coverage of
  the Compton power-law component above 10 keV and the many independent
  observations it provides (typically hundreds); drawbacks are its poor
  low-energy response and spectral resolution.}
\end{figure}

In the CF method, one determines $R_{\rm ISCO}$ by modeling the X-ray
continuum spectrum of the dominant thermal component using {\sc kerrbb2}
(refs.\ 9,28), which is an elaboration of the 1973 model of Novikov \&
Thorne$^{29}$.  The observables are X-ray flux, temperature, distance
$D$, inclination $i$, and BH mass $M$.  In order to obtain reliable
values of $a_*$, it is essential to (1) select X-ray spectra that have a
strong thermal component and (2) have accurate estimates of $M$, $i$ and
$D$, which are typically derived by modeling optical data$^{4}$.

The CF method has delivered the spins of six stellar BHs$^{7-11}$.
Meanwhile, spins for four more BHs are in the works and a half-dozen
more are targeted for future study.  Here we highlight results for three
BH binaries: M33 X-7 (see Fig.\ 1), LMC X-1 (see Fig.\ 2) and GRS
1915+105 (ref.\ 9).  The BH primary of the third system -- a microquasar
with unique and striking properties -- is a near-extreme Kerr BH with a
lower limit on its dimensionless spin parameter of $a_* > 0.98$.  As
illustrated in Figures 7--14 in ref.\ 9, this result is robust in the
sense that it is independent of the details of the data analysis and
insensitive to the uncertainties in mass and distance of the BH.  A
proviso is that one select data of low to moderate luminosity ($L/L_{\rm
Edd} < 0.3$), corresponding to accretion disks that are geometrically
thin ($H/R < 0.1$).

In the Fe K method, one determines $R_{\rm ISCO}$ by modeling the
profile of the broad, skewed line that is formed in the inner disk by
Doppler effects, light bending, and gravitational redshift$^{30}$.  Of
central importance is the effect of the redshift on the red wing of the
line.  This wing extends to very low energies for a rapidly rotating BH
($a_* \sim 1$) because in this case gas orbits near the event horizon.
Relative to the CF method, measuring the extent of this red wing in
order to infer $a_*$ is hindered by the faintness of the signal and
uncertainties in subtracting the continuum.  However, the Fe~K method
has the virtues that it is independent of $M$ and $D$, while the blue
wing of the line even allows an estimate of $i$.  {\it What makes the
Fe~K method enormously important is that it is currently the only viable
approach to measuring the spins of supermassive BHs in AGN.}  (For
further details on the Fe K method, see the white paper by J.\ Miller et
al.).

\vspace{-5mm}

\subsection{High Frequency QPOs and X-ray Polarimetry}

\paragraph{High Frequency QPOs -- Central Question: What is the correct
  model of these strong-field X-ray oscillations?}

Arguably, High Frequency QPOs (HFQPOs; 100--450 Hz) are likely to offer
the most reliable and precise measurement of spin once the correct model
is known.  HFQPOs have been detected in seven BH sources$^{2}$.  They
are of special interest because their frequencies are in the expected
range for matter in orbit near the ISCO.  Four of the seven sources
exhibit harmonic pairs of frequencies in a 3:2 ratio.  These frequencies
(single or pairs) do not vary significantly despite sizable changes in
X-ray luminosity.  Overall, these oscillations appear to be a stable and
identifying feature of a BH that are dependent on the mass and spin of
the BH.  HFQPOs are transient and subtle with typical QPO amplitudes of
$\sim1$\%.  The entire sample of HFQPOs detected by {\it RXTE} at
$>4\sigma$ are illustrated in the white paper by J.\ Tomsick et al.\ and
in ref.\ 2.  As this figure shows, and as argued by Tomsick et al.,
these oscillations are near the sensitivity limit of {\it RXTE} -- the
only mission to have detected them -- and a more powerful timing mission
is required in order to explore and exploit them.

\vspace{-5mm}

\paragraph{X-ray Polarimetry -- Central Question: Is the spin of
  a stellar BH aligned with the orbit vector?}

If a BH's spin were to be misaligned from the orbit vector, the inner
and X-ray-emitting portion of its accretion disk would be warped away
from the outer disk$^{31}$.  One can measure any misalignment by
comparing the orbital inclination angle $i_{\rm orb}$ (routinely
measured to a few degrees via optical observations) to the inclination
of the inner disk $i_{\rm disk}$.  The direct approach to measuring
$i_{\rm disk}$ is via polarimetric studies of BH binaries in
disk-dominated states$^{32,33}$.  The predicted degree of polarization
varies from 0\% to $\sim5$\% as the disk inclination changes from
face-on to edge-on.  Meanwhile, based on two current mission concepts,
even a NASA SMEX-class payload is capable of determining the
polarization of the inner disk to an accuracy of 0.1\% by observing a
typical bright BH binary for about 10 days, thereby constraining the
disk inclination to within a degree or two$^{33}$.

The CF method of measuring BH spin (\S2.1) is straightforward to apply,
the required data are readily obtainable, and even the theory of disk
accretion in strong gravity is tractable (see \S3).  However, the method
is called into question by a single assumption, namely that $i_{\rm
disk} = i_{\rm orb}$.  Unfortunately, the CF method cannot fit for
$i_{\rm disk}$ and check for disk warp because there is a degeneracy
between the inclination and spin parameters$^{33}$.  {\it Therefore,
X-ray polarimetry studies are required to validate this key assumption
of the CF method.}

The Fe K method of measuring spin (\S2.1) will also be greatly
strengthened by polarimetric data.  These data will check on the
inclination estimates obtained via fits to the Fe~K line, or allow this
parameter to be fixed in the fits.  Furthermore, polarimetry can provide
qualitatively new information on source geometry and magnetic fields on
spatial scales comparable to a BH event horizon$^{32,33}$.  This
capability promises to be crucial in defining the geometry of the
coronal source that powers the Fe K line via fluorescence, a source that
is now vaguely and variously characterized as a sphere or a slab or a
lamp post.

Finally, we note that it may be possible to determine BH spin solely via
polarimetry, thereby securing an additional independent measurement of
spin.  This possibility is implied by the pioneering work of Connors et
al.$^{34}$ and has been explored in recent work$^{32,33}$.

\vspace{-5mm}

\section{The Required Key Advances in Observation and Theory in Priority
Order}


\paragraph{An X-ray timing/spectral mission dedicated to the study of bright
Galactic sources:} 

Since its launch in late 1995, {\it RXTE} has revolutionized our
knowledge of stellar BHs and neutron stars because of its large area,
its high data-rate capabilities, and especially because it is a
dedicated observatory that allows sustained, synoptic observations of
these complex and variable systems.  Additionally, its All-Sky Monitor
maintains continual surveillance of the entire sky, which is critically
important because 95\% of the known Galactic stellar BHs are transient
X-ray sources.  A new follow-on mission with order-of-magnitude
improvements in collecting area, data rate, and spectral resolution --
and with sensitivity well below the 3 keV cutoff of {\it RXTE} -- is
required in order to make the next step.  Such a mission will at once
serve the three prime spin methodologies: Briefly, large area is needed
for studying HFQPOs and good spectral resolution for resolving the Fe K
line, while sensitivity down to $\sim1$~keV profits the CF method by
capturing the full thermal continuum spectrum.  For more on such a
mission, see the white paper by Tomsick et al. and ref.\ 35.

In addition to a dedicated mission, it is crucial that the International
X-ray Observatory (IXO) also have the capability to observe bright
Galactic BH transients as targets of opportunity.  Specifically, the
inclusion of the High Time Resolution Spectrometer will insure that no
Rosetta-stone transient slips away unobserved.

\vspace{-5mm}
\paragraph{Advances in computational astrophysics:}

Both the CF and Fe K methods of measuring BH spin assume that disk
radiation cuts off at the ISCO. This is a valid assumption provided that
the accreting gas has negligible torque at the ISCO.  But does the
torque really become small at the ISCO?  The only way to find out is by
means of 3D MHD simulations of the accreting gas in the Kerr metric of a
spinning BH.  This nascent area of research is currently poised to take
off.  Powerful GRMHD codes have been developed and tested$^{36,37}$ and
have begun to provide the first direct estimates of the stress profile
in disks of various thicknesses around non-spinning and spinning
BHs$^{27,38,39}$.  However, the energy dissipation profile and the
corresponding radiative properties of the disk -- the most important
quantities for applying theoretical models to observations -- are still
unknown and require much more work.  The physics of HFQPOs too is likely
to be understood only when GRMHD simulations that include radiation are
carried out.  These developments require (1) numerical GRMHD codes that
can efficiently model thermodynamics and radiation physics and (2)
larger computational resources than are presently available.  The former
can be enabled with adequate funding of theoretical and numerical
research and the latter with serious investment in computer hardware.

\vspace{-5mm}
\paragraph{X-ray polarimetry mission:} 

Most effective would be a modest mission dedicated to observing bright
stellar BHs and neutron stars.  Either of two instrument concepts
developed within the severe constraints of a NASA SMEX-class payload
would be quite effective -- a photoelectron-track polarimeter or a
Bragg-crystal instrument.  Either instrument can, for example, detect
polarization in a 1 Crab source at the $\sim0.3$\% level in 1 day, and
at the $\sim0.1$\% level in 10 days$^{33}$.

\vspace{-5mm}

\section{Goals: 2010--2020}

\begin{itemize}

\item
{Firmly establish the fledgling enterprise of measuring BH spin via the
CF and Fe K methods: Obtain precise and accurate values of spin for
10--20 BHs using one of the methods, and for several BHs using both
methods.}

\item \vspace{-2mm}

{Obtain {\it complete} descriptions of many stellar BHs in order to test
models of jets, GRBs, supernovae, BH formation, BH binary evolution,
etc.}

\item \vspace{-2mm}

{Establish the Fe K methodology for application via IXO to supermassive
BHs.}

\item \vspace{-2mm} 

{Identify the correct model of HFQPOs and so open a third channel for
measuring spin.}

\item \vspace{-2mm}

{Pursue X-ray polarimetry as a means of securing the continuum-fitting
and Fe K methods, and also as a possible fourth avenue to spin.}

\item \vspace{-2mm}

{Develop and test realistic GRHMD models of thin disks in strong gravity.}

\end{itemize}

\noindent {\bf References}

\vspace{1mm}

\footnotesize
\noindent 1.~~Ginzburg, V.~L. 1985, Physics \& Astrophysics: A
Selection of Key Problem (Pergamon) \newline
2.~~Remillard, R.~A., \& McClintock, J.~E. 2006, ARAA, 44, 49
\newline
3.~~Narayan, R., \& McClintock, J.~E. 2008, New Astron.\ Rev.\ 51, 733
   \newline
4.~~Orosz, J.~A., McClintock, J.~E., Narayan, R., et al. 2007, Nature,
  449, 872 \newline
5.~~Prestwich, A.~H., Kilgard, R., Crowther, P.~A., et al.\ 2007, ApJL,
   669, 21 \newline
6.~~Shapiro, S.~L., \& Teukolsky, S. A. 1983, Black Holes, White Dwarfs \&
   Neutron Stars (Wiley) \newline
7.~~Shafee, R., McClintock, J.~E., Narayan, R., et al. 2006, ApJ, 636,
  L113 \newline
8.~~Davis, S.~W., Done, C., \& Blaes, O.~M. 2006, ApJ, 647, 525 \newline
9.~~McClintock, J.~E., Shafee, R., Narayan, R., et al. 2006, ApJ, 652,
  518 \newline
10.~Liu, J., McClintock, J.~E., Narayan, R., et al. 2008, ApJ, 679, L37
  \newline
11.~Gou, L., McClintock, J.~E., Liu, J., et al. 2009, ApJ, submitted,
  arXiv:0901.0920 \newline
12.~Miller, J.~M., Reynolds, C.~S., Fabian, A.~C., et al.\ 2008, ApJ,
  679, L113 \newline
13.~Reis, R.~C., Fabian, A.~C., Ross, R.~R., et al.\ 2008, MNRAS, 387,
   1489 \newline
14.~Reis, R.~C., Fabian, A.~C., Ross, R.~R., \& Miller, J.~M. 2009,
   MNRAS, in press, arXiv:0902.1745 \newline
15.~Brenneman, L.~W., \& Reynolds, C.~S.\ 2006, ApJ, 652, 1028 \newline
16.~Miniutti, G., Fabian, A.~C., Anabuki, N., et al.\ 2007, PASJ, 59,
   315 \newline
17.~Blandford, R.~D., \& Znajek, R.~L. 1977, MNRAS, 179, 433 \newline
18.~Woosley, S.~E. 1993, ApJ, 405, 273 \newline
19.~Fryer, C.~L., \& Kalogera, V. 2001, ApJ, 554, 548 \newline
20.~Portegies Zwart, S.~F., Verbunt, F., \& Ergma, E. 1997, A\&A, 321,
   207 \newline
21.~Campanelli, M., Lousto, C.~O., \& Zlochower, Y. 2006, Phys.\ Rev.\ D, 74,
   041501 \newline
22.~Volonteri, M., Madau, P., Quataert, E., \& Rees, M.~J. 2005, ApJ,
   620, 69 \newline
23.~Zhang, S.~N., Cui, W., \& Chen, W.\ 1997, ApJ, 482, L155 \newline
24.~Tanaka, Y., Nandra, K., \& Fabian, A.~C. et al.\ 1995, Nature, 375,
659 \newline
25.~McClintock, J.~E., et al. 2009, to appear in Black Holes, ed.\ M.\
   Livio (CUP), arXiv:0707.4492 \newline
26.~Reynolds, C.~S., \& Fabian, A.~C. 2008, ApJ, 675, 1048 \newline
27.~Shafee, R., McKinney, J.~C., Narayan, R., et al.\ 2008, ApJ, 687,
   L25 \newline
28.~Li, L.-X., Zimmerman, E.~R., Narayan, R., \& McClintock, J.~E. 2005,
   ApJS, 157, 335 \newline
29.~Novikov, I.~D., \& Thorne, K.~S.\ 1973, in Black Holes, ed.\ C.\
   DeWitt \& B.\ DeWitt (Gordon \& Breach) \newline
30.~Reynolds, C.~S., \& Nowak, M.~A. 2003, Phys.\ Rep., 377, 389 \newline
31.~Fragile, P.~C., Blaes, O.~M., Anninos, P., \& Salmonson, J.~D. 2007,
   ApJ, 668, 417
   \newline
32.~Dov\v{c}iak, M., Muleri, F., \& Goosmann, R.~W. 2008, MNRAS, 391, 32
   \newline
33.~Li, L.-X., Narayan, R., \& McClintock, J.~E. 2009, ApJ, 691, 847
   \newline
34.~Connors, P.~A., Stark, R.~F., \& Piran, T. 1980, ApJ, 235, 224
   \newline
35.~Chakrabarty, D., Ray, P.~S., \& Strohmayer, T.~E. 2008, to appear in
an AIP Conf.\ Proc., arXiv:0809.4029 \newline
36.~Gammie, C.~F., McKinney, J.~C., \& T\'oth, G. 2003, ApJ, 589, 444
   \newline
37.~De Villiers, J.-P., \& Hawley, J.~F. 2003, ApJ, 589, 458 \newline
38.~Krolik, J.~H., Hawley, J.~F., \& Hirose, S. 2005, ApJ, 622, 1008
   \newline
39.~Noble, S.~C., Krolik, J.~H., \& Hawley, J.~F. 2009, ApJ, in press,
   arXiv:0808.3140 \newline

\end{document}